\documentclass[showpacs,preprintnumbers,amsmath,amssymb]{revtex4}
\usepackage{graphicx}
\usepackage{dcolumn}
\usepackage{bm}
\usepackage{slashed}
\begin{document}

\title{New predictions on the mass of the $1^{-+}$ light hybrid
meson from QCD sum rules}

\author{Zhuo-Ran Huang$^1$, Hong-Ying Jin$^1$ and Zhu-Feng Zhang$^2$\\
$^1$Zhejiang Institute of Modern Physics, Zhejiang University, Zhejiang Province, P. R. China\\
$^2$Physics Department, Ningbo University, Zhejiang Province, P. R. China
}

\begin{abstract}
We calculate the coefficients of the dimension-8 quark and gluon condensates
in the current-current correlator of $1^{-+}$ light hybrid current
$g\bar{q}(x)\gamma_{\nu}iG_{\mu\nu}(x)q{(x)}$. With inclusion
of these higher-power corrections and updating the input parameters, we re-analyze the mass of the $1^{-+}$
light hybrid meson from Monte-Carlo based QCD sum rules. Considering the possible
violation of factorization of higher dimensional condensates and variation of $\langle g^3G^3\rangle$, we obtain a conservative mass range 1.72--2.60\,GeV, which favors $\pi_{1}(2015)$ as a better hybrid candidate compared with $\pi_{1}(1600)$ and $\pi_{1}(1400)$.
\end{abstract}


\maketitle

\section{Introduction}

Mesons with exotic quantum numbers have long been attractive in hadron
physics, among which are the $J{}^{PC}=1^{-+}$ isovector states $\pi_{1}(1400)$
, $\pi_{1}(1600)$ and $\pi_{1}(2015)$  identified in the experiments \cite{key-2}.
The construction of these states are not quite clear, four-quark states \cite{Zhang:2001sb,Zhang:2004nb,General:2007bk,Chen:2008qw} and hybrid states are most possible explanations.
Theoretical studies via different
methods have shown that  some of these  states can be considered as good
light hybrid candidates. In the bag model, the predicted mass of $1^{-+}$ light
hybrid meson is around 1.5\,GeV \cite{key-3}; the mass from the flux
tube model is found to be in the range 1.7--1.9\,GeV \cite{key-4-1}; the
lattice QCD prediction of $1^{-+}$ mass is 1.9--2.2\,GeV \cite{key-5-1}.
Calculations based on QCD sum rules \cite{key-1-1} have been conducted
by different groups \cite{key-6-1,key-7-1,key-8-1,key-8-2,key-9,key-9-1} to NLO of
$d\leqq 6$ contributions, and the latest versions of
the predicted mass are $1.80\pm0.06$\,GeV in \cite{key-10-1} and
$1.71\pm0.22$\,GeV in \cite{key-11-1}. Although the hybrid explanation
for $\pi_{1}(1600)$ is supported by previous sum rule analyses, the
hybrid assignment of $\pi_{1}(2015)$ is also proposed \cite{key-10-1}. Thus
the calculation of higher power corrections (HPC) of the OPE is interesting
and of value. How and how much the HPC affect the mass prediction
would lead to totally different conclusions.

In this paper, we focus on the mass prediction of the $1^{-+}$
light hybrid meson using QCD sum rule method. We will first present our calculation of the coefficients of dimension-8 condensates and then include these higher dimensional contributions in the numerical analysis. Due to the possible violation of factorization of $d=6$--$8$ condensates and variation of $\langle g^3G^3\rangle$ condensate, we will consider a conservative range of the mass prediction. We shall compare the results in d$\leqq$8 case with those in d$\leqq$6 case to show
the variation of the mass prediction with inclusion of dimension-8 contributions. In order to obtain an objective conclusion, we shall pay special attention to the fixing of the continuum threshold $s_{0}$, which is not rigorously constrained in the original SVZ sum rules and therefore cause uncertainties. To solve the problem, some authors use the stability criterion to fix $s_{0}$ \cite{key-8-2,key-10-1}. In this work, we shall fit the sum rules following the matching procedure introduced by Leinweber in \cite{Leinweber:1995fn} and successfully performed in some other works \cite{Lee:1996dc,Lee:1997ix,Lee:1997ne,Wang:2008vg}, from which the continuum threshold $s_{0}$ is an output parameter and an uncertainty analysis can be provided.
For the explicit consideration of higher power corrections is not seen very often in previous sum
rule calculations, we will give a slightly more detailed presentation
of our calculation and analysis.

\section{OPE for the current-current correlator}

We start from the two-point correlator
\begin{eqnarray}
\Pi_{\mu\nu}(q^{2}) & = & i\int d^{4}xe^{iqx}\left\langle 0\left|T\left[j_{\mu}(x)j_{\nu}^{+}(0)\right]\right|0\right\rangle \label{eq:1}\\
 & = & (q_{\mu}q_{\nu}-q^{2}g_{\mu\nu})\Pi_{v}(q^{2})+q_{\mu}q_{\nu}\Pi_{s}(q^{2})\nonumber \end{eqnarray}
where $j_{\mu}(x)=g\bar{q}(x)\gamma_{\nu}iG_{\mu\nu}(x)q{(x)}$, and
the invariants $\Pi_{v}(q^{2})$ and $\Pi_{s}(q^{2})$ correspond
respectively to $1^{-+}$ and $0^{++}$ contributions.

The correlator obeys the standard dispersion relation
\begin{equation}
\Pi_{v/s}(q^{2})=\frac{1}{\pi}\int_{0}^{\infty}ds\frac{\textrm{Im}\Pi_{v/s}(s)}{s-q^{2}-i\epsilon}.
\label{eq:20}
\end{equation}

In this paper, we focus on the dimension-8 corrections to the $1^{-+}$
mass. Before showing the higher power results we need to mention that coefficients of dimension-8
quark-related operators of the $1^{-+}$ light hybrid two-point correlator
have been calculated in \cite{key-6-1} and \cite{key-7-1}. In \cite{key-6-1}
there is only a factorized form of the total result and a complete
result is given in \cite{key-7-1}. We obtain a new complete result
which is consistent with the former factorized form but different
from the latter one.

As for dimension-8 gluon operators, there arise IR divergences in the calculation
of the quark loops as the result of setting $m_q=0$ before calculating the
integrals. These IR divergences can be canceled after taking operator mixing
into account. This process can partly check
the calculation about dimension-8 quark and gluon operators and modify the
finite part of the coefficients of gluon condensates. Some good examples
for the case of $\bar q q$ scalar and vector currents are given in \cite{key-14,key-15}.

According to the numbers of quark operators in the condensates, dimension-8
quark condensates can be classified into two groups: two-quark $d=8$
condensates and four-quark $d=8$ condensates. Only the formers can be
mixed to $d=8$ gluon condensates in LO. We use the dimensional regularization
in $n=4-\epsilon$ space-time dimensions, thus the $O(\epsilon)$ terms
of the two-quark $d=8$ condensates can be obtained, which are needed
to be multiplied by the $\frac{1}{\epsilon}$ subtractions
to modify the finite part of the quark loop calculations (see Eq.\eqref{eq:16}).

The dimension-8 quark contributions (corresponding to Feynman diagrams
in Figure~\ref{fig:1}) are listed in Appendix A.

\begin{figure}[htbp]
\centering
\includegraphics[scale=0.5]{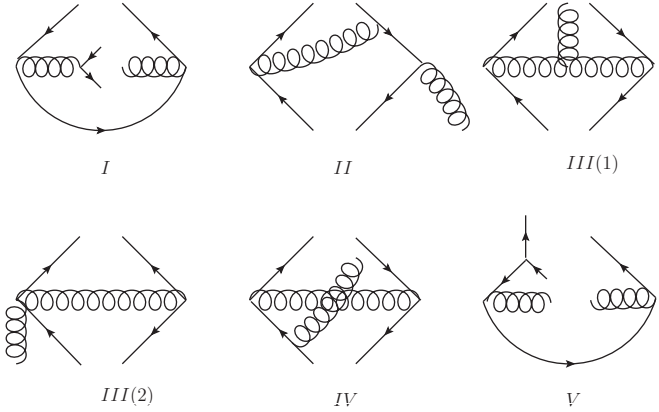}
\caption{\label{fig:1}Feynman diagrams of dimension-8 quark contributions.}
\end{figure}

Dimension-8 contributions of gluon condensates come from the calculations
of quark loops. Here we give the quark propagator up to term $O(q^{-5})$
needed in the calculation of the quark loops:
\begin{eqnarray}
S(q) & = & S_{0}(q)+\frac{ig}{2}G_{\rho\mu}S_{0}(q)\gamma_{\mu}\frac{\partial}{\partial q_{\rho}}S_{0}(q)+\frac{g}{3}D_{\alpha}G_{\rho\mu}S_{0}(q)\gamma_{\mu}\frac{\partial}{\partial q_{\alpha}}\frac{\partial}{\partial q_{\rho}}S_{0}(q)\label{eq:8}\\
 & - & \frac{ig}{8}D_{\alpha1}D_{\alpha2}G_{\rho\mu}S_{0}(q)\gamma_{\mu}\frac{\partial}{\partial q_{\alpha1}}\frac{\partial}{\partial q_{\alpha2}}\frac{\partial}{\partial q_{\rho}}S_{0}(q)-\frac{g^{2}}{4}G_{\rho\mu}G_{\sigma\nu}S_{0}(q)\gamma_{\mu}\frac{\partial}{\partial q_{\rho}}\left[S_{0}(q)\gamma_{\nu}\frac{\partial}{\partial q_{\sigma}}S_{0}(q)\right],\nonumber \end{eqnarray}
where $D_{\mu}=\partial_{\mu}-igA_{\mu}$ and $S_{0}(q)=\frac{1}{\slashed q}$.

For a massless quark Eq.\eqref{eq:8} can be rewritten as
\begin{eqnarray}
S(q) & = & \frac{\slashed q}{q^{2}}+\frac{1}{q^{4}}gq_{\alpha}\tilde{G}_
{\alpha\beta}\gamma_{\beta}\gamma_{5}\label{eq:9}\\
 & + & \frac{1}{q^{6}}\left[-\frac{2}{3}g\left(q_{\alpha}q_{\rho}D_{\rho}
 G_{\alpha\beta}\gamma_{\beta}-J_{\mu}q_{\mu}\slashed q+q^{2}\slashed J\right)+2igq_{\alpha}q_{\rho}D_{\rho}\tilde{G}_{\alpha\beta}\gamma_{\beta}\gamma_{5}\right]\nonumber \\
 & + & \frac{1}{q^{8}}\{-2igq_{\gamma}D_{\gamma}\left(q^{2}\slashed
 J-q_{\mu}J_{\mu}\slashed q\right)+\left[-4g\left(q_{\gamma}D_{\gamma}\right)^{2}
 +gq^{2}D^{2}\right]q_{\alpha}\tilde{G}_{\alpha\beta}\gamma_{\beta}\gamma_{5}+
 2ig\left(q_{\gamma}D_{\gamma}\right)^{2}q_{\alpha}G_{\mu\alpha}\gamma_{\mu}\nonumber \\
 & + & 2g^{2}q_{\mu}q_{\alpha}G_{\mu\rho}G_{\alpha\rho}\slashed q+2g^{2}q^{2}
 q_{\mu}G_{\alpha\rho}G_{\rho\mu}\gamma_{\alpha}+ig^{2}q^{2}q_{\alpha}\left(
 \tilde{G}_{\mu\beta}G_{\alpha\beta}-G_{\mu\beta}\tilde{G}_{\alpha\beta}\right)\gamma_{\mu}\gamma_{5}\},\nonumber \end{eqnarray}
where $\tilde{G}_{\alpha\beta}=\frac{1}{2}\varepsilon_{\alpha\beta\mu\nu}G_{\mu\nu}$,
$\gamma_{5}=-\frac{i}{4}\varepsilon_{\alpha\beta\mu\nu}\gamma_{\alpha}
\gamma_{\beta}\gamma_{\mu}\gamma_{\nu},J_{\mu}=D_{\nu}G_{\mu\nu}=
g\underset{uds}{\sum}\overline{\psi}\gamma_{\mu}T^{a}\psi T^{a}$.
Eq.\eqref{eq:9} can also be seen in \cite{key-16} and \cite{key-15}, but the last
term of \eqref{eq:9} is missed in \cite{key-16} and not consistent
with \cite{key-15}. We use \eqref{eq:8} rather than \eqref{eq:9}
in practical calculations for \eqref{eq:8} is more convenient in program
calculations.

Gluon contributions from calculations of quark loops (the corresponding
Feynman diagrams are depicted in Figure~\ref{fig:2}) are listed in Appendix A.

\begin{figure}[htbp]
\centering
\includegraphics[scale=0.5]{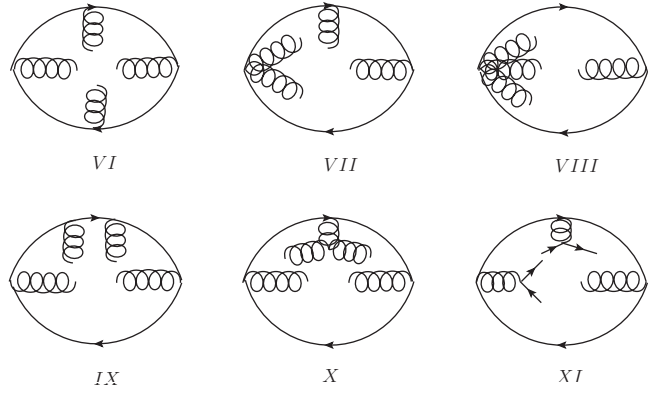}
\caption{\label{fig:2}Feynman diagrams of dimension-8 gluon contributions.}
\end{figure}

\begin{table}[htbp]
\caption{\label{tab:1}The independent $d=8$  two-quark condensates and coefficients. $(\gamma_{\mu}\gamma_{\nu}\gamma_{\rho}\gamma_{\sigma})_{-}$  and
$(\gamma_{\mu}\gamma_{\rho}\gamma_{\sigma})_{-}$ are totally anti-symmetric tensors.}
\begin{minipage}{0.5\textwidth}
\begin{ruledtabular}
\begin{tabular}{cccccc}
j & $Q_{j}$ & $C_{j}^{V}$ & $D_{j}^{V}$ & $C_{j}^{S}$ & $D_{j}^{S}$\\
\hline
1 & $-ig^{2}\bar{q}[\slashed DG_{\mu\nu},G_{\rho\sigma}](\gamma_{\mu}\gamma_{\nu}\gamma_{\rho}\gamma_{\sigma})_{-}q$ & $\frac{1}{72}$ & $\frac{1}{96}$ & $\frac{1}{24}$ & $\frac{5}{288}$\\
2 & $-ig^{2}\bar{q}[\slashed DG_{\mu\nu},G_{\mu\nu}]q$ & $-\frac{1}{36}$ & $-\frac{1}{48}$ & $-\frac{1}{12}$ & $-\frac{5}{144}$\\
3 & $-ig^{2}\bar{q}[G_{\mu\nu},G_{\rho\sigma}]D_{\nu}(\gamma_{\mu}\gamma_{\rho}\gamma_{\sigma})_{-}q$ & $\frac{1}{18}$ & $\frac{1}{24}$ & $\frac{1}{6}$ & $\frac{5}{72}$\\
4 & $-ig^{2}\bar{q}\{G_{\mu\rho},G_{\mu\nu}\}\gamma_{\nu}D_{\rho}q$ & $-\frac{1}{9}$ & $\frac{1}{9}$ & $\frac{2}{3}$ & $\frac{7}{36}$\\
5 & $-ig^{2}\bar{q}\{D_{\mu}G_{\nu\mu},G_{\rho\sigma}\}(\gamma_{\nu}\gamma_{\rho}\gamma_{\sigma})_{-}q$ & $-\frac{1}{9}$ & $-\frac{5}{144}$ & $-\frac{1}{12}$ & $-\frac{1}{18}$\\
6 & $-ig^{2}\bar{q}[D_{\mu}G_{\nu\mu},G_{\nu\alpha}]\gamma_{\alpha}q$ & $-\frac{1}{18}$ & $-\frac{1}{24}$ & $-\frac{1}{6}$ & $-\frac{5}{72}$\\
7 & $-ig^{2}\bar{q}D^{2}D_{\nu}G_{\alpha\nu}\gamma_{\alpha}q$ & 0 & 0 & 0 & 0\\
\end{tabular}
\end{ruledtabular}
\end{minipage}
\end{table}

\begin{table}[htbp]
\caption{\label{tab:2}The mixing coefficients in \eqref{eq:16} \cite{key-14}.}
\begin{minipage}{0.5\textwidth}
\begin{ruledtabular}
\begin{tabular}{ccccccccccccc}
j & 2 & 3 & 3 & 3 & 3 & 3 & 3 & 4 & 4 & 4 & 6 & 7\\
\hline
i & 6 & 1 & 2 & 3 & 4 & 5 & 6 & 1 & 3 & 4 & 5 & 7\\
\hline
$Z_{i}^{j}$ & -6 & 6 & -6 & -12 & 12 & -6 & -3 & -6 & 12 & 12 & 6 & -6\\
\end{tabular}
\end{ruledtabular}
\end{minipage}
\end{table}

The two-quark $d=8$ condensates in \eqref{eq:3} and \eqref{eq:6} can
be expanded in the basis $\left\{ Q_{j}\right\} $ listed in Table~\ref{tab:1},
using the equations of motion and charge conjugation transformation
and setting $m_{q}=0$. And we also list the expanding coefficients in Table~\ref{tab:1}
and the mixing coefficients of quark condensates with gluon condensates
in Table~\ref{tab:2} \cite{key-14}. After taking operator mixing into account, the
corrective terms to gluon contributions are obtained:
\begin{eqnarray}
\Pi_{c}^{G}(q) & = & (q_{\mu}q_{\nu}-q^{2}g_{\mu\nu})\sum_{j=1}^{7}\left(C_{j}^{V}+\epsilon D_{j}^{V}\right)\sum_{i=1}^{7}Z_{i}^{j}g^{n_{i}}O_{i}/\left(72\pi^{2}\omega\right)\frac{1}{q^{4}}\label{eq:16}\\
 & + & q_{\mu}q_{\nu}\sum_{j=1}^{7}\left(C_{j}^{S}+\epsilon D_{j}^{S}\right)\sum_{i=1}^{7}Z_{i}^{j}g^{n_{i}}O_{i}/\left(72\pi^{2}\omega\right)\frac{1}{q^{4}}\nonumber \\
 & = & (q_{\mu}q_{\nu}-q^{2}g_{\mu\nu})[\left(-\frac{5}{864\pi^{2}}+\frac{1}{72\pi^{2}}\frac{1}{\omega}\right)g^{4}O_{1}+\left(-\frac{1}{288\pi^{2}}-\frac{1}{216\pi^{2}}\frac{1}{\omega}\right)g^{4}O_{2}\nonumber \\
 & + & \left(\frac{5}{432\pi^{2}}-\frac{1}{36\pi^{2}}\frac{1}{\omega}\right)g^{4}O_{3}+\left(\frac{11}{432\pi^{2}}-\frac{1}{108\pi^{2}}\frac{1}{\omega}\right)g^{4}O_{4}+\left(-\frac{1}{144\pi^{2}}-\frac{1}{108\pi^{2}}\frac{1}{\omega}\right)g^{3}O_{5}]\frac{1}{q^{4}}\nonumber \\
 & + & q_{\mu}q_{\nu}[\left(-\frac{1}{96\pi^{2}}-\frac{1}{24\pi^{2}}\frac{1}{\omega}\right)g^{4}O_{1}+\left(-\frac{5}{864\pi^{2}}-\frac{1}{72\pi^{2}}\frac{1}{\omega}\right)g^{4}O_{2}\nonumber \\
 & + & \left(\frac{1}{48\pi^{2}}+\frac{1}{12\pi^{2}}\frac{1}{\omega}\right)g^{4}O_{3}+\left(\frac{19}{432\pi^{2}}+\frac{5}{36\pi^{2}}\frac{1}{\omega}\right)g^{4}O_{4}+\left(-\frac{5}{432\pi^{2}}-\frac{1}{36\pi^{2}}\frac{1}{\omega}\right)g^{3}O_{5})]\frac{1}{q^{4}},\nonumber \end{eqnarray}
where $\frac{1}{\omega}=\frac{1}{\epsilon}+\frac{1}{2}\ln4\pi-\frac{\gamma_E}{2}$ and
by which the IR divergences in \eqref{eq:13},\eqref{eq:14} and
\eqref{eq:15} can be canceled. Thus the total dimension-8 contributions
are the sum of \eqref{eq:16} and \eqref{eq:7}$-$\eqref{eq:15}:
\begin{eqnarray}
\Pi_{\mu\nu}^{d=8}(q^{2}) & = & (q_{\mu}q_{\nu}-q^{2}g_{\mu\nu})[-\frac{1}{24}g^{3}\left\langle \bar{q}q\right\rangle \left\langle \bar{q}Gq\right\rangle \label{eq:17}\\
 & + & \left(-\frac{1}{108\pi^{2}}+\frac{1}{144\pi^{2}}\ln\frac{-q^{2}}{\mu^{2}}\right)g^{4}O_{1}+\left(\frac{1}{216\pi^{2}}-\frac{1}{432\pi^{2}}\ln\frac{-q^{2}}{\mu^{2}}\right)g^{4}O_{2}\nonumber \\
 & + & \left(-\frac{1}{108\pi^{2}}-\frac{1}{72\pi^{2}}\ln\frac{-q^{2}}{\mu^{2}}\right)g^{4}O_{3}+\left(\frac{1}{54\pi^{2}}-\frac{1}{216\pi^{2}}\ln\frac{-q^{2}}{\mu^{2}}\right)g^{4}O_{4}\nonumber \\
 & + & \left(\frac{1}{864\pi^{2}}-\frac{1}{216\pi^{2}}\ln\frac{-q^{2}}{\mu^{2}}\right)g^{3}O_{5}+\frac{1}{288\pi^{2}}g^{2}O_{7}+\frac{1}{288\pi^{2}}g^{3}O_{8}]\frac{1}{q^{4}}\nonumber \\
 & + & q_{\mu}q_{\nu}[-\frac{11}{27}g^{3}\left\langle \bar{q}q\right\rangle \left\langle \bar{q}Gq\right\rangle \nonumber \\
 & + & \left(\frac{7}{576\pi^{2}}-\frac{1}{48\pi^{2}}\ln\frac{-q^{2}}{\mu^{2}}\right)g^{4}O_{1}+\left(-\frac{7}{516\pi^{2}}-\frac{1}{144\pi^{2}}\ln\frac{-q^{2}}{\mu^{2}}\right)g^{4}O_{2}\nonumber \\
 & + & \left(-\frac{13}{288\pi^{2}}+\frac{1}{24\pi^{2}}\ln\frac{-q^{2}}{\mu^{2}}\right)g^{4}O_{3}+\left(-\frac{7}{288\pi^{2}}+\frac{5}{72\pi^{2}}\ln\frac{-q^{2}}{\mu^{2}}\right)g^{4}O_{4}\nonumber \\
 & + & \left(\frac{11}{576\pi^{2}}-\frac{1}{72\pi^{2}}\ln\frac{-q^{2}}{\mu^{2}}\right)g^{3}O_{5}+\frac{1}{192\pi^{2}}g^{2}O_{7}]\frac{1}{q^{4}},\nonumber \end{eqnarray}
where $\left\langle \overline{q}Gq\right\rangle =\left\langle \overline{q}\frac{\lambda^{a}}{2}G_{\mu\nu}^{a}\sigma_{\mu\nu}q\right\rangle $, $\sigma_{\mu\nu}=\frac{i}{2}[\gamma_{\mu,}\gamma_{\nu}]$.

Notice that the quark condensates have been factorized in \eqref{eq:17} so as to conduct the sum rule analysis. As is well-known, factorization hypothesis may have large uncertainties as observed in some other channels \cite{Bertlmann:1984ih,Bertlmann:1987ty,Launer:1983ib,Narison:1992ru,Narison:1995jr,Narison:2009vy}. Therefore we shall consider the possible violation of factorization for quark condensates in the numerical analysis.
As for the values of $\left\langle G^{4}\right\rangle $
condensates $\left(O_{1}-O_{4}\right)$, one may also think of using
factorization. However, for reasons in \cite{key-18}, factorization
hypothesis may as well not be reliable in $\left\langle G^{4}\right\rangle $
case. Therefore we choose to use a modified factorization proposed
in \cite{key-19} and supported in \cite{key-13}, which suggests an
overestimation of factorization and are based on two technologies: factorization
of quartic heavy quark condensates and heavy quark expansion. In the framework
of this modified factorization, $O_{1}-O_{4}$ can be expressed in
terms of the condensate $\phi=\textrm{Tr}G_{\nu\mu}G_{\mu\rho}\textrm{Tr}G_{\nu\tau}G_{\tau\rho}$, which
has been clarified in \cite{key-19} to reasonably satisfy the factorization
approximation. Thus after fitting $\phi$ using factorization, $O_{1}-O_{4}$
can be estimated as follows
\begin{eqnarray}
g^{4}O_{1} & = & \frac{1}{12}\left\langle g^{2}G^{2}\right\rangle ^{2},g^{4}O{}_{2}=-\frac{5}{48}\left\langle g^{2}G^{2}\right\rangle ^{2}+2g^{4}\phi,\label{eq:21}\\
g^{4}O_{3} & = & g^{4}O_{4}=-\frac{1}{192}\left\langle g^{2}G^{2}\right\rangle ^{2}+\frac{1}{2}g^{4}\phi.\nonumber \end{eqnarray}

With regard to other $d=8$ gluon condensates, a scale $M^{2}\approx 0.3\,\textrm{GeV}^{2}$
is estimated in \cite{key-13,key-20}, which characterizes the average
off-shellness of the vacuum gluons and quarks:
\begin{equation}
g^{3}O_{5}=-\frac{3}{2}g^{4}\left\langle \bar{q}q\right\rangle ^{2}M^{2},g^{2}O_{7}=-\frac{4}{3}g^{4}\left\langle \bar{q}q\right\rangle ^{2}M^{2},g^{3}O_{8}=\left\langle g^{3}G^{3}\right\rangle M^{2},\label{eq:22}\end{equation}
and we shall also consider the violation of factorization of $O_5$ and $O_7$ in the matching procedure.

\section{QCD Sum Rules for the $1^{-+}$ light hybrid meson}

The $d\leqq6$ contributions to $\Pi_{v}(q^{2})$ including the NLO
corrections to the perturbative and the $\left\langle \alpha_{s}G^{2}\right\rangle $
and $\alpha_{s}\left\langle \overline{q}q\right\rangle ^{2}$ terms
can be found in \cite{key-6-1,key-7-1,key-8-1,key-8-2,key-9,key-9-1}. $\Pi_{v}^{d\leqq6}(q^{2})$ can be written as

\begin{equation}
\Pi_v^{d\leqq6}(q^2)=a_{11} q^4\ln\frac{-q^2}{\mu^2}+a_{12} q^4\ln^2\frac{-q^2}{\mu^2}+b_{11}
\ln\frac{-q^2}{\mu^2}+b_{12}\ln^2\frac{-q^2}{\mu^2}+c_{11}\frac{1}{-q^2}+c_{12}\frac{1}{-q^2}\ln\frac{-q^2}{\mu^2}
\end{equation}
with
\begin{gather*}
a_{11}=-\frac{\alpha_s(\mu)}{240\pi^3}\left(1+\frac{1301}{240}\frac{\alpha_s(\mu)}{\pi}\right), ~~
a_{12}=\frac{\alpha_s(\mu)}{240\pi^3}\frac{17}{72}\frac{\alpha_s(\mu)}{\pi},\\
b_{11}=-\frac{1}{36\pi}\langle \alpha_s G^2\rangle \left(1-\frac{145}{72}\frac{\alpha_s(\mu)}{\pi}\right)-
\frac{2}{9}\frac{\alpha_s(\mu)}{\pi} \langle m_q\bar qq\rangle,\\
b_{12}=-\frac{1}{36\pi}\langle \alpha_s G^2\rangle \frac{8}{9} \frac{\alpha_s(\mu)}{\pi},\\
c_{11}=-\frac{4\pi}{9}k_1 \alpha_s \langle \bar qq\rangle^2\left( 1+\frac{1}{108}\frac{\alpha_s(\mu)}{\pi}\right)
-\frac{1}{192\pi^2} \langle g^3 G^3\rangle,\\
c_{12}=-\frac{4\pi}{9}k_1\alpha_s\langle \bar qq\rangle^2\frac{47}{72}\frac{\alpha_s(\mu)}{\pi},
\end{gather*}
where $\alpha_s(\mu)=4\pi/(9\ln(\mu^2/\Lambda^2_{\textrm{QCD}}))$ is the running coupling constant for three flavors, and $k_1$ indicates the deviation from vacuum saturation of $d=6$ quark condensates.

In addition, $\Pi_v^{d=8}(q^2)$ can be obtained from \eqref{eq:17},\eqref{eq:21} and \eqref{eq:22}:
\begin{equation}
\Pi_v^{d=8}(q^2)=d_{11}\frac{1}{q^4}+d_{12}\frac{1}{q^4}\ln\frac{-q^2}{\mu^2}
\end{equation}
with
\begin{gather*}
d_{11}=-\frac{\pi}{6}k_2\alpha_s(\mu)\langle\bar qq\rangle \langle g\bar qGq\rangle-\frac{1}{216}\langle
\alpha_s G^2\rangle^2-\frac{11}{108}k_2\alpha_s(\mu)\cdot \alpha_s\langle\bar qq\rangle^2\cdot M^2+\frac{1}{288\pi^2}\langle g^3G^3\rangle M^2,\\
d_{12}=-\frac{1}{648}\langle \alpha_s G^2\rangle^2+\frac{1}{9}k_2\alpha_s(\mu)\cdot \alpha_s\langle \bar qq\rangle^2\cdot M^2,
\end{gather*}
where $k_2$ indicates the deviation from vacuum saturation of d=8 condensates.

The Borel transformation of $\Pi^{\textrm{OPE}}_{v}(q^2)$ can be written as
\begin{equation}
\label{eq:ope}
\begin{split}
\Pi^{\textrm{OPE}}_v(\tau)\equiv&\frac{1}{\tau}\hat B_\tau\Pi^{\textrm{OPE}}_{v}(q^2)= a_{11}\frac{-2}{\tau^3}
+a_{12}\frac{2}{\tau^3} (2 \gamma_E-3+2\ln(\tau \mu^2))+b_{11}\frac{-1}{\tau}+b_{12}\frac{2}{\tau}(\gamma_E+\ln(\tau\mu^2))\\
&+c_{11}+c_{12}(-\gamma_E-\ln(\tau\mu^2))
+d_{11}\tau+d_{12}\tau(1-\gamma_E-\ln(\tau\mu^2)).
\end{split}
\end{equation}

By using the single narrow resonance spectral density ansatz
$\textrm{Im}\Pi_v^{\textrm{phen}}(s)=\pi f_H^2 m_H^4 \delta(s-m_H^2)+\textrm{Im}\Pi_v^{\textrm{OPE}}(s)\theta(s-s_0)$, where $s_0$
is the continuum threshold, $f_H$ and $m_H$ denote the coupling of the hadron to the current and the mass of the hadron
respectively, we can obtain the phenomenological representation of $\Pi_v^{\textrm{phen}}(\tau,s_0,f_H,m_H)$ via the dispersion relation:
\begin{equation}
\label{eq:phen}
\Pi_v^{\textrm{phen}}(\tau,s_0,f_H,m_H)=\frac{1}{\pi}\int_0^\infty{\rm Im}\Pi_v^{\textrm{phen}}(s)e^{-s\tau}ds.
\end{equation}
Then the master equation for QCDSR can be written as
\begin{equation}
\label{eq:qcdsr}
\Pi_v^{\textrm{OPE}}(\tau)=\Pi_v^{\textrm{phen}}(\tau,s_0,f_H,m_H),
\end{equation}
physical properties of the relevant hadron, i.e., $m_H$, $f_H$ and $s_0$,
should satisfy Eq.\eqref{eq:qcdsr}.

In order to present the influence of the $d=8$ contributions, we will
conduct the sum rule analysis both in $d\leqq6$ and $d\leqq8$ cases. Before
those, we should clarify our criteria for establishing the sum rule
window in which the mass prediction is reliable.
On the OPE side, we wish the Borel parameter $\tau$ is as small
as possible so that power series converge as quickly as possible.
On the hadron spectrum side, our wish is the opposite,
because a larger $\tau$ can better suppress contributions of
the excited states and continuum.
The common procedure without considering the higher power
contributions is usually as follows: 1.keep the highest
dimensional contributions (HDC, normally dimension-6 contributions)
no more than 10\% (or 15\%) of the total
OPE contributions to ensure the convergence of  OPE, which gives
the upper bound of $\tau$; 2.make sure that the contributions
from the continuum are under 50\% of the total contributions, which
ensures the validity of the narrow resonance ansatz and gives the
lower bound of $\tau$. For our case, if we require dimension-8
contributions are less than 15 percent, it means we choose a window
with a larger upper bound compared with $d\leqq$6 case. This choice
enhances suppression of excited states and continuum, but the
convergence of OPE gets worse, which increases the uncertainties
of the OPE side. On the other hand, if we still require the
dimension-6 contributions are less than 15 percent, uncertainties
from the truncation of OPE are indeed decreased(because the
dimension-8 contributions are now taken into account), but the
validity of the narrow resonance ansatz is not improved.
Apparently, to keep a balance should be a good resolution.
Our choice is that make sure both $1\%< d=8$  contributions $<5\%$
and $20\%< d=6$ contributions $<35\%$ (correspondingly the
perturbative and $d<6$ contributions would be totally $120\%$ --$140\%$
because the signs of the $d=6$ and $d=8$ contributions are minus),
which ensures the OPE series converge in a proper trend and also
a larger upper bound of $\tau$ is obtained compared with $d \leqq$6
case,thus uncertainties from both sides of the master equation are reduced.

In the original SVZ sum rules, the continuum threshold $s_{0}$ cannot be
rigorously constrained. To overcome this shortcoming and make our conclusion
more reliable, we use a weighted-least-square method following Leinweber
\cite{Leinweber:1995fn} to match the two sides of Eq.\eqref{eq:qcdsr} in the sum rule window.

By randomly generating 200 sets of Gaussian distributed phenomenological input parameters with given uncertainties (10\%
uncertainties, which are typical uncertainties in QCDSR) at
$\tau_j=\tau_\textrm{min}+(\tau_\textrm{max}-\tau_\textrm{min})\times(j-1)/(n_B-1)$, where $n_B=21$, we can estimate the
standard deviation $\sigma_{\textrm{OPE}}(\tau_j)$ for $\Pi_v^{\textrm{OPE}}(\tau_j)$. Then,the phenomenological output parameters
$s_0$, $f_H$ and $m_H$ can be obtained by minimizing
\begin{equation}
\chi^2=\sum_{j=1}^{n_B}\frac{(\Pi^{\textrm{OPE}}(\tau_j)-\Pi^{\textrm{phen}}(\tau_j,s_0,f_H,m_H))^2}{\sigma_{\textrm{OPE}}^2(\tau_j)}.
\end{equation}

We use two sets of parameters as the central values of inputs (see Table~\ref{tab:input})
to conduct the matching procedures respectively. Values in set I are from a recent review of
QCD sum rules \cite{key-21}. We choose this set of values to avoid subjective factors in choosing the inputs.
We also notice that the value of $g^3\langle G^3\rangle$ in
\cite{key-21} is different from the previous one used in \cite{key-9-1,key-10-1,key-11-1}. This value changes from $1.2 \,\textrm{GeV}^2\langle \alpha_{s}G^2\rangle$ (from dilute gas instantons \cite{Novikov:1981xi} and lattice calculations \cite{D'Elia:1997ne}) to $8.2 \,\textrm{GeV}^2\langle\alpha_{s}G^2\rangle$ (from charmonium systems \cite{key-13}),
which largely affects the mass predictions. To make our conclusions more reliable and to provide a comparison of the $d\leqq6$ results in this work and those from previous analyses, we maintain the value of
$g^3\langle G^3\rangle$ the small one in set II.

As in our previous paper \cite{key-11-1}, we generate 2000 sets of Gaussian distributed input parameters with 10\% uncertainties, and for each set
we minimize $\chi^2$ to obtain a set of phenomenological output parameters, after this procedure is finished,
we can estimate the uncertainties of $s_0$, $f_H$ and $m_H$.

\begin{table}[htbp]
\caption{\label{tab:input} Different input phenomenological parameters (at scale $\mu_0=1$\,GeV).}
\begin{ruledtabular}
\begin{tabular}{cccccccc}
 & $\Lambda_{\textrm{QCD}}$ & $\langle \alpha_s G^2\rangle$ &  $m_q$ & $\langle g^3 G^3\rangle$ & $\alpha_s\langle\bar qq\rangle^2$ & $\langle g\bar qGq\rangle$\\
\hline
Set I & $0.353\,\textrm{GeV}$ & $0.07\,\textrm{GeV}^4$& $0.007\,\textrm{GeV}$ &$8.2\,\textrm{GeV}^2 \langle \alpha_s G^2\rangle$ & $1.5\times10^{-4}\,\textrm{GeV}^4$ & $0.8\,\textrm{GeV}^2 \langle \bar qq\rangle$\\
Set II & $0.353\,\textrm{GeV}$ & $0.07\,\textrm{GeV}^4$& $0.007\,\textrm{GeV}$ &$1.2\,\textrm{GeV}^2 \langle \alpha_s G^2\rangle$ & $1.5\times10^{-4}\,\textrm{GeV}^4$ & $0.8\,\textrm{GeV}^2 \langle \bar qq\rangle$\\
\end{tabular}
\end{ruledtabular}
\end{table}

Finally, before proceeding with numerical calculations, renormalization-group (RG) improvement of the sum rules, i.e.,
substitutions $\mu^2\to1/\tau$ in Eq.\eqref{eq:qcdsr}, is needed \cite{Narison:1981ts}. In addition, the anomalous dimensions
for condensate $\langle g^3G^3\rangle$ and $\langle \bar qq\rangle \langle g\bar qGq\rangle$  also should
be implemented by multiplying  $\langle g^3G^3\rangle$ and $\langle \bar qq\rangle \langle g\bar qGq\rangle$
by a factor $L(\mu_0)^{-23/27}$ and  $L(\mu_0)^{10/27}$ respectively, where $L(\mu_0)=[\ln(1/(\tau\Lambda^2_{\textrm{QCD}}))/\ln(\mu^2_0/\Lambda^2_{\textrm{QCD}})]$,
 $\mu_0$ is the renormalization scale for condensates\cite{key-1-1,book2}.
The coupling constant $f_H$ also should be multiplied by a factor $L(m)^{-32/81}$,
$f_H$ then receives its value at hybrid mass shell. In this paper, we neglect the
anomalous dimensions for operators $O_1- O_8$, which are not calculated yet and very likely to have small effects on the mass prediction.

Our matching results with input parameters in Set I and Set II can be seen in Apendix B. We consider violation of factorization by different factors (up to 3 for dimension-6 condensates \cite{Bertlmann:1984ih,Bertlmann:1987ty,Launer:1983ib,Narison:1992ru,Narison:1995jr,Narison:2009vy}, and up to 5 for dimension-8 condensates \cite{Narison:2004vz}). The upper bounds of sum rule windows in each table are obtained by different demands on $|$HDC$|$/OPE. The matching results, including the medians and the asymmetric standard deviations from the medians for $s_0$, $m_H$ and $f^2_H$, are reported. By inputting Gaussian distributed input parameters with 10\% uncertainties, we obtain some Gaussian-like distribution results for $s_0$, $m_H$ and $f_H^2$ with uncertainties $<$10\%, this implies the matching results are very stable with different input parameters. Following our criteria above for establishing the window, the phenomenological outputs in the fourth column of each table are the most reliable (optimal windows) for each case. In fact, we can see that the predictions are not very sensitive to the variation of the range of the window. All output parameters slightly decrease for stronger constraints on contributions from HDC. In addition, we also list the results deduced from $d\leqq6$ contributions in the optimal windows of $d\leqq8$ cases to show the variations of the sum rules in these regions after considering the dimension-8 contributions.

Under the considerations of possible violation of factorization and different values of $\langle g^3G^3\rangle$, we obtain a mass range 1.88--2.60\,GeV from the optimal windows. Furthermore, We shall also consider effects of tachyonic gluon mass \cite{Chetyrkin:1998yr,Narison:2001ix,Narison:2009ag} beyond the original OPE as in \cite{key-10-1}. The lowest order correction due to this effect can be found in \cite{key-10-1}, which leads to decreases in hybrid mass predictions. Taking this effect into account, the lower bound of the mass range would further decrease, therefore we obtain as conclusion a quite conservative range of the predicted mass, i.e. 1.72--2.60\,GeV, which covers $\pi_{1}(2015)$ and is hard to favor $\pi_{1}(1400)$ and $\pi_{1}(1600)$ as hybrids.

As a supplement of our analysis, we also consider as above a conservative mass range in $d\leqq6$ case. With the small $\langle g^3G^3\rangle$ value used in \cite{key-9-1,key-10-1}, the range is 1.55--2.29\,GeV, which is consistent with previous predictions within errors and covers $\pi_{1}(1600)$. In the large $\langle g^3G^3\rangle$ case, the predicted range is 1.84--2.46\,GeV. Notice that even in this case, the hybrid assignment of $\pi_{1}(1600)$ can hardly be favored.

More details of the weighted-least-square matching method can be seen in our previous work on the $1^{-+}$ light hybrid meson \cite{key-11-1}. In that work, we concentrate ourselves on the sum rule analysis based on the matching procedure, especially the uncertainty analysis. However,the dimension-8 coefficients used there are not a complete form (only the factorized quark condensates in \cite{key-6-1}). And we follow our earlier works \cite{key-9,key-9-1} in choosing the inputs there and neglect the violation of saturation hypothesis. Moreover, the sum rule window there is just established by keeping HDC$<$10\% as the common procedure, lacking in an explicit consideration of the convergence of OPE. All these lead to the discrepancy of the predictions.

\section{Summary}

We have calculated the dimension-8 coefficients of the two-point
correlator of the current $g\bar{q}(x)\gamma_{\nu}iG_{\mu\nu}(x)q{(x)}$. We find
that the inclusion of the dimension-8 condensate contributions in QCDSR analysis increases
the predicted mass, and so does the effect of violation of factorization of higher dimensional condensates.
Besides, the variation of the value of  $\langle g^3G^3\rangle$ also have effects on increasing the mass prediction.
Therefore all these new effects suggest that the $1^{-+}$ light hybrid meson may have a larger mass compared with previous QCDSR predictions.
From our analysis, the conservative range of the mass is 1.72--2.60\,GeV, which covers $\pi_{1}(2015)$ and disfavors the hybrid explanations for
$\pi_{1}(1600)$ and $\pi_{1}(1400)$. One can also consider the central value 2.16\,GeV in this range as a very crude estimation of the mass.

As for the effect of the dimension-8 contributions in determining the $1^{-+}$ mass, it's hard to draw a definite conclusion due to the uncertainties
from violation of factorization. From the data in Appendix B, we find that 4\%--9\% underestimation would be led to by neglecting the $d=8$ condensate contributions in the case
of the $1^{-+}$ light hybrid state.

\begin{acknowledgments}
This work is supported by NSFC under grant 11175153, 11205093 and 11347020, and supported by K. C. Wong Magna Fund in Ningbo University.
\end{acknowledgments}

\clearpage

\begin{appendix}

\section*{Appendix A: Results of Calculations of Feynman Diagrams}

We list in this appendix the results of the calculations of the Feynman
diagrams in Figure~\ref{fig:1} and Figure~\ref{fig:2}.

\begin{eqnarray}
\pi_{\mu\nu}^{\textrm{I}}(q) & = & (q_{\mu}q_{\nu}-q^{2}g_{\mu\nu})[\left(-\frac{1}{18}-\frac{\epsilon}{24}\right)ig^{2}\left\langle \bar{q}D_{\mu}G_{\mu\alpha}G_{\nu\beta}\gamma_{\alpha}\gamma_{\nu}\gamma_{\beta}q\right\rangle \label{eq:3}\\
 & + & \left(-\frac{2}{9}+\frac{\epsilon}{36}\right)ig^{2}\left\langle \bar{q}D_{\rho}G_{\mu\alpha}G_{\mu\beta}\gamma_{\alpha}\gamma_{\rho}\gamma_{\beta}q\right\rangle +\left(-\frac{1}{18}-\frac{\epsilon}{24}\right)ig^{2}\left\langle \bar{q}D_{\nu}G_{\mu\alpha}G_{\nu\beta}\gamma_{\alpha}\gamma_{\mu}\gamma_{\beta}q\right\rangle ]\frac{1}{q^{4}}\nonumber \\
 & + & q_{\mu}q_{\nu}[\left(-\frac{1}{6}-\frac{5\epsilon}{72}\right)ig^{2}\left\langle \bar{q}D_{\mu}G_{\mu\alpha}G_{\nu\beta}\gamma_{\alpha}\gamma_{\nu}\gamma_{\beta}q\right\rangle +\left(\frac{1}{3}+\frac{\epsilon}{18}\right)ig^{2}\left\langle \bar{q}D_{\rho}G_{\mu\alpha}G_{\mu\beta}\gamma_{\alpha}\gamma_{\rho}\gamma_{\beta}q\right\rangle \nonumber \\
 & + & \left(-\frac{1}{6}-\frac{5\epsilon}{72}\right)ig^{2}\left\langle \bar{q}D_{\nu}G_{\mu\alpha}G_{\nu\beta}\gamma_{\alpha}\gamma_{\mu}\gamma_{\beta}q\right\rangle ]\frac{1}{q^{4}},\nonumber \end{eqnarray}

\begin{eqnarray}
\pi_{\mu\nu}^{\textrm{II}}(q) & = & (q_{\mu}q_{\nu}-q^{2}g_{\mu\nu})[\frac{5}{18}ig^{3}\left\langle \bar{q}\gamma_{\alpha}T^{a}q\bar{q}T^{a}G_{\mu\beta}\gamma_{\mu}\gamma_{\alpha}\gamma_{\beta}q\right\rangle \label{eq:4}\\
 & - & \frac{1}{18}ig^{3}\left\langle \bar{q}\gamma_{\alpha}T^{a}q\bar{q}T^{a}G_{\mu\beta}\gamma_{\alpha}\gamma_{\mu}\gamma_{\beta}q\right\rangle -\frac{2}{9}ig^{3}\left\langle \bar{q}\gamma_{\alpha}T^{a}q\bar{q}T^{a}G_{\alpha\beta}\gamma_{\beta}q\right\rangle ]\frac{1}{q^{4}}\nonumber \\
 & + & q_{\mu}q_{\nu}[-\frac{1}{6}ig^{3}\left\langle \bar{q}\gamma_{\alpha}T^{a}q\bar{q}T^{a}G_{\mu\beta}\gamma_{\mu}\gamma_{\alpha}\gamma_{\beta}q\right\rangle \nonumber \\
 & + & \frac{5}{6}ig^{3}\left\langle \bar{q}\gamma_{\alpha}T^{a}q\bar{q}T^{a}G_{\mu\beta}\gamma_{\alpha}\gamma_{\mu}\gamma_{\beta}q\right\rangle -\frac{2}{3}ig^{3}\left\langle \bar{q}\gamma_{\alpha}T^{a}q\bar{q}T^{a}G_{\alpha\beta}\gamma_{\beta}q\right\rangle ]\frac{1}{q^{4}},\nonumber \end{eqnarray}

\begin{eqnarray}
\pi_{\mu\nu}^{\textrm{III}}(q) & = & (q_{\mu}q_{\nu}-q^{2}g_{\mu\nu})\left(\frac{1}{12}g^{3}\left\langle f^{abc}G_{\alpha\beta}^{a}\bar{q}\gamma_{\alpha}T^{b}q\bar{q}T^{c}\gamma_{\beta}q\right\rangle \right)\frac{1}{q^{4}}\\
 & + & q_{\mu}q_{\nu}\left(-\frac{3}{4}g^{3}\left\langle f^{abc}G_{\alpha\beta}^{a}\bar{q}\gamma_{\alpha}T^{b}q\bar{q}T^{c}\gamma_{\beta}q\right\rangle \right)\frac{1}{q^{4}},\nonumber \end{eqnarray}

\begin{eqnarray}
\pi_{\mu\nu}^{\textrm{{IV}}}(q) & = & (q_{\mu}q_{\nu}-q^{2}g_{\mu\nu})[\frac{1}{18}g^{3}\left\langle \bar{q}G_{\mu\nu}T^{a}\sigma_{\mu\nu}\gamma_{\alpha}q\bar{q}T^{a}\gamma_{\alpha}q\right\rangle \label{eq:5}\\
 & + & \frac{1}{36}ig^{3}\left\langle \bar{q}\gamma_{\beta}T^{a}q\bar{q}G_{\alpha\beta}T^{a}\gamma_{\alpha}q\right\rangle +\frac{1}{18}g^{3}\left\langle \bar{q}T^{a}G_{\mu\nu}\gamma_{\alpha}\sigma_{\mu\nu}q\bar{q}T^{a}\gamma_{\alpha}q\right\rangle \nonumber \\
 & - & \frac{1}{36}ig^{3}\left\langle \bar{q}\gamma_{\beta}T^{a}q\bar{q}T^{a}G_{\alpha\beta}\gamma_{\alpha}q\right\rangle +\frac{2}{9}g^{2}\left\langle \bar{q}\overleftarrow{D}_{\alpha}T^{a}\gamma_{\beta}\overrightarrow{D}_{\alpha}q\bar{q}T^{a}\gamma_{\beta}q\right\rangle ]\frac{1}{q^{4}}\nonumber \\
 & + & q_{\mu}q_{\nu}[\frac{1}{4}g^{3}\left\langle \bar{q}G_{\mu\nu}T^{a}\sigma_{\mu\nu}\gamma_{\alpha}q\bar{q}T^{a}\gamma_{\alpha}q\right\rangle -\frac{1}{4}ig^{3}\left\langle \bar{q}\gamma_{\beta}T^{a}q\bar{q}G_{\alpha\beta}T^{a}\gamma_{\alpha}q\right\rangle \nonumber \\
 & + & \frac{1}{4}g^{3}\left\langle \bar{q}T^{a}G_{\mu\nu}\gamma_{\alpha}\sigma_{\mu\nu}q\bar{q}T^{a}\gamma_{\alpha}q\right\rangle +\frac{1}{4}ig^{3}\left\langle \bar{q}\gamma_{\beta}T^{a}q\bar{q}T^{a}G_{\alpha\beta}\gamma_{\alpha}q\right\rangle \nonumber \\
 & + & g^{2}\left\langle \bar{q}\overleftarrow{D}_{\alpha}T^{a}\gamma_{\beta}\overrightarrow{D}_{\alpha}q\bar{q}T^{a}\gamma_{\beta}q\right\rangle ]\frac{1}{q^{4}},\nonumber \end{eqnarray}

\begin{eqnarray}
\pi_{\mu\nu}^{\text{\mbox{v}}}(q) & = & (q_{\mu}q_{\nu}-q^{2}g_{\mu\nu})[\left(-\frac{2}{9}+\frac{\epsilon}{36}\right)ig^{2}\left\langle \bar{q}\overleftarrow{D}_{\rho}G_{\mu\alpha}G_{\mu\beta}\gamma_{\alpha}\gamma_{\varrho}\gamma_{\beta}q\right\rangle \label{eq:6}\\
 & + & \left(-\frac{1}{18}-\frac{\epsilon}{24}\right)ig^{2}\left\langle \bar{q}\overleftarrow{D}_{\mu}G_{\mu\alpha}G_{\nu\beta}\gamma_{\alpha}\gamma_{\nu}\gamma_{\beta}q\right\rangle +\left(-\frac{1}{18}-\frac{\epsilon}{24}\right)ig^{2}\left\langle \bar{q}\overleftarrow{D}_{\nu}G_{\mu\alpha}G_{\nu\beta}\gamma_{\alpha}\gamma_{\mu}\gamma_{\beta}q\right\rangle ]\frac{1}{q^{4}}\nonumber \\
 & + & q_{\mu}q_{\nu}[\left(\frac{1}{3}+\frac{\epsilon}{18}\right)ig^{2}\left\langle \bar{q}\overleftarrow{D}_{\rho}G_{\mu\alpha}G_{\mu\beta}\gamma_{\alpha}\gamma_{\varrho}\gamma_{\beta}q\right\rangle +\left(-\frac{1}{6}-\frac{5\epsilon}{72}\right)ig^{2}\left\langle \bar{q}\overleftarrow{D}_{\mu}G_{\mu\alpha}G_{\nu\beta}\gamma_{\alpha}\gamma_{\nu}\gamma_{\beta}q\right\rangle \nonumber \\
 & + & \left(-\frac{1}{6}-\frac{5\epsilon}{72}\right)ig^{2}\left\langle \bar{q}\overleftarrow{D}_{\nu}G_{\mu\alpha}G_{\nu\beta}\gamma_{\alpha}\gamma_{\mu}\gamma_{\beta}q\right\rangle ]\frac{1}{q^{4}},\nonumber \end{eqnarray}

where $\sigma_{\mu\nu}=\frac{i}{2}[\gamma_{\mu},\gamma_{\nu}]$, and
the total result can be factorized as follows

\begin{equation}
\Pi_{q}^{d=8}(q^{2})=(q_{\mu}q_{\nu}-q^{2}g_{\mu\nu})(-\frac{1}{24}g^{3}\left\langle \bar{q}q\right\rangle \left\langle \bar{q}Gq\right\rangle \frac{1}{q^{4}})+q_{\mu}q_{\nu}(-\frac{11}{27}g^{3}\left\langle \bar{q}q\right\rangle \left\langle \bar{q}Gq\right\rangle \frac{1}{q^{4}}),\label{eq:7}\end{equation}

which is consistent with the factorized form in \cite{key-6-1}(the
condensate $\left\langle \bar{q}D_{\rho}G_{\mu\alpha}G_{\mu\beta}\gamma_{\alpha}\gamma_{\rho}\gamma_{\beta}q\right\rangle $ that
cannot be factorized are set to $-\frac{7ig}{72}\left\langle \bar{q}q\right\rangle \left\langle \bar{q}Gq\right\rangle $
based on the formula $\left\langle \bar{q}D_{\rho}G_{\mu\alpha}G_{\mu\beta}\gamma_{\alpha}\gamma_{\rho}\gamma_{\beta}q\right\rangle =-\frac{7ig}{72}\left\langle \bar{q}q\right\rangle \left\langle \bar{q}Gq\right\rangle -\frac{1}{4}\left\langle \bar{q}d^{abc}D_{\mu}(G_{\alpha\rho}^{a}G_{\mu\beta}^{b})T^{c}\gamma_{\alpha}\gamma_{\rho}\gamma_{\beta}q\right\rangle $+gluon
condensates ).

\begin{eqnarray}
\pi_{\mu\nu}^{\text{\mbox{VI}}}(q) & = & (q_{\mu}q_{\nu}-q^{2}g_{\mu\nu})\label{eq:10}\\
 & \times & [-\frac{1}{144\pi^{2}}g^{4}O_{1}-\frac{1}{144\pi^{2}}g^{4}O_{2}-\frac{1}{144\pi^{2}}g^{4}O_{3}+\frac{1}{16\pi^{2}}g^{4}O_{4}]\frac{1}{q^{4}}\nonumber \\
 & + & q_{\mu}q_{\nu}[-\frac{1}{48\pi^{2}}g^{4}O_{1}-\frac{1}{48\pi^{2}}g^{4}O_{2}-\frac{1}{48\pi^{2}}g^{4}O_{3}+\frac{1}{16\pi^{2}}g^{4}O_{4}]\frac{1}{q^{4}},\nonumber \end{eqnarray}

\begin{eqnarray}
\pi_{\mu\nu}^{\text{\mbox{VII}}}(q) & = & (q_{\mu}q_{\nu}-q^{2}g_{\mu\nu})\label{eq:11}\\
 & \times & [-\frac{1}{288\pi^{2}}g^{4}O_{1}+\frac{1}{288\pi^{2}}g^{4}O_{2}-\frac{1}{144\pi^{2}}g^{4}O_{3}+\frac{1}{144\pi^{2}}g^{4}O_{4}+\frac{7}{576\pi^{2}}g^{3}O_{8}]\frac{1}{q^{4}}\nonumber \\
 & + & q_{\mu}q_{\nu}[\frac{1}{96\pi^{2}}g^{4}O_{1}-\frac{1}{96\pi^{2}}g^{4}O_{2}-\frac{1}{16\pi^{2}}g^{4}O_{3}+\frac{1}{16\pi^{2}}g^{4}O_{4}+\frac{1}{192\pi^{2}}g^{3}O_{8}]\frac{1}{q^{4}},\nonumber \end{eqnarray}

\begin{eqnarray}
\pi_{\mu\nu}^{\text{\mbox{VIII}}}(q) & = & (q_{\mu}q_{\nu}-q^{2}g_{\mu\nu})\label{eq:12}\\
 & \times & [-\frac{1}{288\pi^{2}}g^{4}O_{1}+\frac{1}{288\pi^{2}}g^{4}O_{2}+\frac{1}{54\pi^{2}}g^{4}O_{3}-\frac{1}{54\pi^{2}}g^{4}O_{4}\nonumber \\
 & - & \frac{1}{864\pi^{2}}g^{3}O_{5}+\frac{1}{288\pi^{2}}g^{2}O_{7}-\frac{5}{864\pi^{2}}g^{3}O_{8}]\frac{1}{q^{4}}\nonumber \\
 & + & q_{\mu}q_{\nu}[\frac{1}{576\pi^{2}}g^{4}O_{1}-\frac{1}{576\pi^{2}}g^{4}O_{2}+\frac{7}{288\pi^{2}}g^{4}O_{3}-\frac{7}{288\pi^{2}}g^{4}O_{4}\nonumber \\
 & - & \frac{1}{576\pi^{2}}g^{3}O_{5}+\frac{1}{192\pi^{2}}g^{2}O_{7}-\frac{1}{288\pi^{2}}g^{3}O_{8}]\frac{1}{q^{4}},\nonumber \end{eqnarray}

\begin{eqnarray}
\pi_{\mu\nu}^{\text{\mbox{IX}}}(q) & = & (q_{\mu}q_{\nu}-q^{2}g_{\mu\nu})\label{eq:13}\\
 & \times & [\left(-\frac{1}{72\pi^{2}}\frac{1}{\hat{\epsilon}}+\frac{11}{864\pi^{2}}\right)g^{4}O_{1}+\left(\frac{1}{216\pi^{2}}\frac{1}{\hat{\epsilon}}+\frac{5}{864\pi^{2}}\right)g^{4}O_{2}\nonumber \\
 & + & \left(\frac{5}{72\pi^{2}}\frac{1}{\hat{\epsilon}}-\frac{19}{864\pi^{2}}\right)g^{4}O_{3}+\left(-\frac{7}{216\pi^{2}}\frac{1}{\hat{\epsilon}}-\frac{53}{864\pi^{2}}\right)g^{4}O_{4}]\frac{1}{q^{4}}\nonumber \\
 & + & q_{\mu}q_{\nu}[\left(\frac{1}{24\pi^{2}}\frac{1}{\hat{\epsilon}}+\frac{13}{288\pi^{2}}\right)g^{4}O_{1}+\left(\frac{1}{72\pi^{2}}\frac{1}{\hat{\epsilon}}+\frac{11}{864\pi^{2}}\right)g^{4}O_{2}\nonumber \\
 & + & \left(-\frac{1}{12\pi^{2}}\frac{1}{\hat{\epsilon}}+\frac{1}{72\pi^{2}}\right)g^{4}O_{3}+\left(-\frac{5}{36\pi^{2}}\frac{1}{\hat{\epsilon}}-\frac{41}{216\pi^{2}}\right)g^{4}O_{4}]\frac{1}{q^{4}},\nonumber \end{eqnarray}

\begin{eqnarray}
\pi_{\mu\nu}^{\text{\mbox{X}}}(q) & = & (q_{\mu}q_{\nu}-q^{2}g_{\mu\nu})\label{eq:14}\\
 & \times & [\left(\frac{1}{72\pi^{2}}\frac{1}{\hat{\epsilon}}-\frac{1}{288\pi^{2}}\right)g^{4}O_{3}+\left(-\frac{1}{72\pi^{2}}\frac{1}{\hat{\epsilon}}+\frac{1}{288\pi^{2}}\right)g^{4}O_{4}+\left(-\frac{1}{144\pi^{2}}\frac{1}{\hat{\epsilon}}+\frac{1}{576\pi^{2}}\right)g^{3}O_{8}]\frac{1}{q^{4}}\nonumber \\
 & + & q_{\mu}q_{\nu}\left(\frac{1}{48\pi^{2}}g^{4}O_{3}-\frac{1}{48\pi^{2}}g^{4}O_{4}-\frac{1}{96\pi^{2}}g^{3}O_{8}\right)\frac{1}{q^{4}},\nonumber \end{eqnarray}

\begin{eqnarray}
\pi_{\mu\nu}^{\text{\mbox{XI}}}(q) & = & (q_{\mu}q_{\nu}-q^{2}g_{\mu\nu})\label{eq:15}\\
 & \times & [-\frac{1}{432\pi^{2}}g^{4}O_{1}+\frac{1}{432\pi^{2}}g^{4}O_{2}-\frac{1}{18\pi^{2}}\frac{1}{\hat{\epsilon}}g^{4}O_{3}+\frac{1}{18\pi^{2}}\frac{1}{\hat{\epsilon}}g^{4}O_{4}\nonumber \\
 & + & \left(\frac{1}{108\pi^{2}}\frac{1}{\hat{\epsilon}}+\frac{1}{108\pi^{2}}\right)g^{3}O_{5}+\left(\frac{1}{144\pi^{2}}\frac{1}{\hat{\epsilon}}-\frac{1}{216\pi^{2}}\right)g^{3}O_{8}]\frac{1}{q^{4}}\nonumber \\
 & + & q_{\mu}q_{\nu}[-\frac{1}{72\pi^{2}}g^{4}O_{1}+\frac{1}{72\pi^{2}}g^{4}O_{2}-\frac{1}{24\pi^{2}}g^{4}O_{3}\nonumber \\
 & + & \frac{1}{24\pi^{2}}g^{4}O_{4}+\left(\frac{1}{36\pi^{2}}\frac{1}{\hat{\epsilon}}+\frac{7}{216\pi^{2}}\right)g^{3}O_{5}+\frac{5}{576\pi^{2}}g^{3}O_{8}]\frac{1}{q^{4}},\nonumber \end{eqnarray}
where $O_{1}=\textrm{Tr}\left(G_{\mu\nu}G_{\mu\nu}G_{\alpha\beta}G_{\alpha\beta}\right)$, $O_{2}=\textrm{Tr}\left(G_{\mu\nu}G_{\alpha\beta}G_{\mu\nu}G_{\alpha\beta}\right)$, $O_{3}=\textrm{Tr}\left(G_{\mu\nu}G_{\upsilon\alpha}G_{\alpha\beta}G_{\beta\mu}\right)$,
$O_{4}=\textrm{Tr}\left(G_{\mu\nu}G_{\alpha\beta}G_{\nu\alpha}G_{\beta\mu}\right)$, $O_{5}=f_{abc}G_{\mu\nu}^{a}j_{\mu}^{b}j_{\nu}^{c}$, $O_{6}=f_{abc}G_{\mu\nu}^{a}j_{\lambda}^{b}D_{\lambda}G_{\mu\nu}^{c}$, $O_{7}=j_{\mu}^{a}D^{2}j_{\mu}^{a}$,
$O_{8}=f^{abc}G_{\mu\nu}^{a}G_{\nu\lambda}^{b}D^{2}G_{\lambda\mu}^{c}$,
and $\frac{1}{\hat{\epsilon}}=\frac{1}{\epsilon}-\frac{1}{2}\ln\frac{-q^{2}}{\mu^{2}}$+$\frac{1}{2}\ln4\pi$-$\frac{\gamma_E}{2}$.

\section*{Appendix B: Results of Numerical Analysis}

\begin{table}[htbp]
\centering
\begin{tabular}{|c|c|c|c|c|c|}
\hline
$|$HDC$|$/OPE & $<$10\% & $<$5\% & $<$1\% & $<$15\% (d$\leqq$6)& - (d$\leqq$6)\\
\hline
  $[\tau_{\text{min}},\tau_{\text{max}}]$/GeV$^{-2}$ & [0.24,0.89] & [0.26,0.81]  & [0.32,0.60] & [0.36, 0.45]& [0.32,0.60]\\
  \hline
  $s_0$/GeV$^2$ & $10.73^{+0.78}_{-0.65}$ & $9.70^{+0.63}_{-0.55}$  & $7.92^{+0.39}_{-0.37}$ & $7.01^{+0.35}_{-0.34}$ & $7.64^{+0.45}_{-0.42}$ \\
  \hline
   $m_H$/GeV & $2.34^{+0.06}_{-0.05}$ & $2.29^{+0.05}_{-0.05}$  & $2.16^{+0.05}_{-0.04}$ & $2.08^{+0.06}_{-0.06}$ & $2.13^{+0.05}_{-0.05}$ \\
   \hline
   $f_H^2$/$10^{-3}$GeV$^2$ & $0.80^{+0.05}_{-0.05}$ & $0.74^{+0.05}_{-0.05}$  & $0.63^{+0.04}_{-0.04}$ &
  $0.57^{+0.04}_{-0.03}$ & $0.62^{+0.04}_{-0.04}$\\
  \hline
\end{tabular}
\caption{\label{tab:result1a} Matching results with input parameters in Set I (10\% uncertainties for input phenomenological parameters, $k_1=k_2=1$).}
\end{table}

\begin{table}[htbp]
\centering
\begin{tabular}{|c|c|c|c|c|c|}
\hline
$|$HDC$|$/OPE & $<$10\% & $<$5\% & $<$2\% & $<$15\% (d$\leqq$6)& - (d$\leqq$6)\\
\hline
  $[\tau_{\text{min}},\tau_{\text{max}}]$/GeV$^{-2}$ & [0.23,0.69] & [0.25,0.61]  & [0.28,0.51] & [0.32, 0.38]& [0.28,0.51]\\
  \hline  $s_0$/GeV$^2$ & $11.47^{+0.88}_{-0.74}$ & $10.40^{+0.69}_{-0.61}$  & $9.38^{+0.52}_{-0.48}$ & $8.10^{+0.43}_{-0.40}$ & $8.75^{+0.56}_{-0.50}$ \\
  \hline
   $m_H$/GeV & $2.50^{+0.07}_{-0.06}$ & $2.43^{+0.06}_{-0.06}$  & $2.36^{+0.06}_{-0.06}$ & $2.25^{+0.06}_{-0.06}$ & $2.30^{+0.06}_{-0.06}$ \\
   \hline
   $f_H^2$/$10^{-3}$GeV$^2$ & $0.78^{+0.04}_{-0.04}$ & $0.72^{+0.04}_{-0.04}$  & $0.66^{+0.04}_{-0.04}$ &
  $0.58^{+0.04}_{-0.03}$ & $0.62^{+0.04}_{-0.03}$\\
  \hline
\end{tabular}
\caption{\label{tab:result1b} Matching results with input parameters in Set I (10\% uncertainties for input phenomenological parameters, $k_1=k_2=2$).}
\end{table}

\begin{table}[htbp]
\centering
\begin{tabular}{|c|c|c|c|c|c|}
\hline
$|$HDC$|$/OPE & $<$10\% & $<$5\%  & $<$2\% & $<$15\% (d$\leqq$6) &-(d$\leqq$6)\\
\hline
  $[\tau_{\text{min}},\tau_{\text{max}}]$/GeV$^{-2}$ & [0.21,0.60] & [0.23,0.53]  & [0.25,0.44] & [0.29, 0.34]& [0.25,0.44]\\
  \hline
  $s_0$/GeV$^2$ & $12.66^{+1.07}_{-0.87}$ & $11.54^{+0.85}_{-0.72}$  & $10.45^{+0.64}_{-0.57}$ & $9.12^{+0.53}_{-0.49}$ & $9.65^{+0.65}_{-0.58}$ \\
  \hline
   $m_H$/GeV & $2.65^{+0.08}_{-0.07}$ & $2.58^{+0.07}_{-0.07}$ & $2.51^{+0.07}_{-0.06}$ & $2.39^{+0.07}_{-0.06}$ & $2.44^{+0.07}_{-0.07}$ \\
   \hline
   $f_H^2$/$10^{-3}$GeV$^2$ & $0.79^{+0.04}_{-0.04}$ & $0.73^{+0.04}_{-0.04}$ & $0.67^{+0.04}_{-0.04}$ &
  $0.60^{+0.03}_{-0.03}$ & $0.63^{+0.03}_{-0.03}$\\
  \hline
\end{tabular}
\caption{\label{tab:result1c} Matching results with input parameters in Set I (10\% uncertainties for input phenomenological parameters, $k_1=k_2=3$).}
\end{table}

\begin{table}[htbp]
\centering
\begin{tabular}{|c|c|c|c|}
\hline
$|$HDC$|$/OPE & $<$10\% & $<$5\%  & $<$3\%\\
\hline
  $[\tau_{\text{min}},\tau_{\text{max}}]$/GeV$^{-2}$ & [0.21,0.53] & [0.23,0.46]  & [0.25,0.41]\\
  \hline
  $s_0$/GeV$^2$ & $12.23^{+0.85}_{-0.73}$ & $11.17^{+0.67}_{-0.60}$  & $10.66^{+0.58}_{-0.53}$ \\
  \hline
   $m_H$/GeV & $2.64^{+0.07}_{-0.07}$ & $2.58^{+0.06}_{-0.06}$ & $2.54^{+0.06}_{-0.06}$\\
   \hline
   $f_H^2$/$10^{-3}$GeV$^2$ & $0.76^{+0.04}_{-0.04}$ & $0.71^{+0.04}_{-0.04}$ & $0.68^{+0.04}_{-0.04}$ \\
  \hline
\end{tabular}
\caption{\label{tab:result1d} Matching results with input parameters in Set I (10\% uncertainties for input phenomenological parameters, $k_1=3$, $k_2=5$).}
\end{table}

\begin{table}[htbp]
\centering
\begin{tabular}{|c|c|c|c|c|c|}
\hline
$|$HDC$|$/OPE & $<$10\% & $<$7\% & $<$5\% &  $<$15\% (d$\leqq$6)& - (d$\leqq$6)\\
\hline
  $[\tau_{\text{min}},\tau_{\text{max}}]$/GeV$^{-2}$ & [0.36,0.81] & [0.38,0.74] & [0.39,0.68] & [0.48,0.59] & [0.39,0.68]\\
  \hline
  $s_0$/GeV$^2$ & $6.98^{+0.37}_{-0.40}$ &  $6.62^{+0.33}_{-0.36}$ & $6.32^{+0.30}_{-0.33}$ &   $5.12^{+0.28}_{-0.33}$ & $5.17^{+0.29}_{-0.35}$ \\
  \hline
   $m_H$/GeV & $1.98^{+0.04}_{-0.05}$ &  $1.95^{+0.04}_{-0.05}$ & $1.93^{+0.04}_{-0.05}$ & $1.77^{+0.04}_{-0.05}$ & $1.78^{+0.04}_{-0.05}$ \\
   \hline
   $f_H^2$/$10^{-3}$GeV$^2$ & $0.65^{+0.05}_{-0.04}$ &  $0.62^{+0.05}_{-0.04}$ & $0.60^{+0.04}_{-0.04}$ &
  $0.53^{+0.04}_{-0.03}$ & $0.54^{+0.04}_{-0.04}$\\
  \hline
\end{tabular}
\caption{\label{tab:result2a} Matching results with input parameters in Set II (10\% uncertainties for input phenomenological parameters, $k_1=k_2=1$).}
\end{table}

\begin{table}[htbp]
\centering
\begin{tabular}{|c|c|c|c|c|c|}
\hline
$|$HDC$|$/OPE & $<$10\% & $<$7\% & $<$4\% &  $<$15\% (d$\leqq$6)& - (d$\leqq$6)\\
\hline
  $[\tau_{\text{min}},\tau_{\text{max}}]$/GeV$^{-2}$ & [0.29,0.66] & [0.30,0.61] & [0.32,0.54] & [0.39,0.45] & [0.32,0.54]\\
  \hline
  $s_0$/GeV$^2$ & $8.85^{+0.57}_{-0.53}$ & $8.41^{+0.50}_{-0.47}$ &  $7.97^{+0.43}_{-0.42}$ &  $6.57^{+0.41}_{-0.40}$ & $6.70^{+0.43}_{-0.42}$ \\
  \hline
   $m_H$/GeV & $2.25^{+0.06}_{-0.06}$ & $2.22^{+0.05}_{-0.06}$ & $2.18^{+0.05}_{-0.05}$ &  $2.02^{+0.06}_{-0.06}$ & $2.04^{+0.06}_{-0.06}$ \\
   \hline
   $f_H^2$/$10^{-3}$GeV$^2$ & $0.67^{+0.04}_{-0.04}$ & $0.65^{+0.04}_{-0.04}$ &  $0.62^{+0.04}_{-0.04}$ &
  $0.54^{+0.04}_{-0.03}$ & $0.55^{+0.04}_{-0.03}$\\
  \hline
\end{tabular}
\caption{\label{tab:result2b} Matching results with input parameters in Set II (10\% uncertainties for input phenomenological parameters, $k_1=k_2=2$).}
\end{table}

\begin{table}[htbp]
\centering
\begin{tabular}{|c|c|c|c|c|c|}
\hline
$|$HDC$|$/OPE & $<$10\% & $<$7\% & $<$4\% &  $<$15\% (d$\leqq$6)& - (d$\leqq$6)\\
\hline
  $[\tau_{\text{min}},\tau_{\text{max}}]$/GeV$^{-2}$ & [0.25,0.59] & [0.26,0.54] & [0.27,0.48] & [0.33,0.38] & [0.27,0.48] \\
  \hline
  $s_0$/GeV$^2$ & $10.57^{+0.82}_{-0.72}$ & $10.01^{+0.70}_{-0.64}$ &  $9.41^{+0.59}_{-0.55}$ &  $7.82^{+0.51}_{-0.50}$ & $8.09^{+0.58}_{-0.55}$\\
  \hline
   $m_H$/GeV & $2.46^{+0.07}_{-0.07}$ & $2.42^{+0.07}_{-0.07}$ & $2.38^{+0.06}_{-0.07}$ &  $2.22^{+0.07}_{-0.07}$ & $2.24^{+0.07}_{-0.07}$ \\
   \hline
   $f_H^2$/$10^{-3}$GeV$^2$ & $0.71^{+0.05}_{-0.04}$ & $0.68^{+0.04}_{-0.04}$ &  $0.65^{+0.04}_{-0.04}$ &
  $0.56^{+0.04}_{-0.03}$ & $0.58^{+0.04}_{-0.03}$\\
  \hline
\end{tabular}
\caption{\label{tab:result2c} Matching results with input parameters in Set II (10\% uncertainties for input phenomenological parameters, $k_1=k_2=3$).}
\end{table}

\begin{table}[htbp]
\centering
\begin{tabular}{|c|c|c|c|}
\hline
$|$HDC$|$/OPE & $<$10\% & $<$7\% & $<$5\% \\
\hline
  $[\tau_{\text{min}},\tau_{\text{max}}]$/GeV$^{-2}$ & [0.24,0.53] & [0.25,0.49] & [0.26,0.45]\\
  \hline
  $s_0$/GeV$^2$ & $10.59^{+0.69}_{-0.63}$ & $10.15^{+0.62}_{-0.57}$ &  $9.76^{+0.55}_{-0.52}$\\
  \hline
   $m_H$/GeV & $2.49^{+0.07}_{-0.07}$ & $2.46^{+0.06}_{-0.06}$ & $2.43^{+0.06}_{-0.06}$ \\
   \hline
   $f_H^2$/$10^{-3}$GeV$^2$ & $0.71^{+0.04}_{-0.04}$ & $0.68^{+0.04}_{-0.04}$ &  $0.66^{+0.04}_{-0.04}$\\
  \hline
\end{tabular}
\caption{\label{tab:result2d} Matching results with input parameters in Set II (10\% uncertainties for input phenomenological parameters, $k_1=3$, $k_2=5$).}
\end{table}

\end{appendix}

\end{document}